# Enhancing Quasi-Two-Dimensional Electronic Structure in $Bi_2O_2Se$ via Chlorine Substitution

Yong-Zhen Jiang[1*], Qi-Xun Wen[2*], Xiang Xu[1*], Su-Tao Sun[2*], Feng-Fan Ren[1], Zi-Xuan Zhou[1], Xing-Yuan You[2], Yu-Cheng Kan[2], Zheng Zhang[1], Jie-Yi Liu[3], Yu-Zhou Zhao[1], Zheng-Tai Liu[4,5], Mao Ye[4,5], Shu-Hua Yao[2], Jian Zhou[2], Yan-Bin Chen[2], Zhong-Kai Liu[1], Yu-Lin Chen[1], Yang-Yang Lv[2†], Cheng Chen[1†]

*[1]State Key Laboratory of Quantum Functional Materials and ShanghaiTech Laboratory for Topological Physics, School of Physical Science and Technology, ShanghaiTech University, Shanghai,201210, China*

*[2] National Laboratory of Solid State Microstructures, Department of Materials Science and Engineering, Department of Physics, Nanjing University, Nanjing, 210093, China*

*[3]Diamond Light Source, Harwell Science and Innovation Campus, Didcot OX11 0DE, UK*

*[4]Shanghai Synchrotron Radiation Facility, Shanghai Advanced Research Institute, Chinese Academy of Sciences, Shanghai, 201210, China*

*[5]National Key Laboratory of Materials for Integrated Circuits, Shanghai Institute of Microsystem and Information Technology, Chinese Academy of Sciences, Shanghai 200050, China*

**These authors contributed equally to the work*

*† Correspondence to: lvyangws0801@nju.edu.cn, chencheng1@shanghaitech.edu.cn*

## Abstract:

**Bismuth oxyselenide ($Bi_2O_2Se$) has emerged as a leading semiconductor candidate for next-generation electronics and quantum phenomena research. However, its strong ionic interlayer bonding of Se layers fundamentally restricts the full potential of this otherwise promising quasi-two-dimensional (2D) material. In this work, we demonstrate chlorine substitution as an effective strategy to introduce van der Waals gaps and modify the electronic structure of the system. Through comprehensive transport and angle-resolved photoemission spectroscopy (ARPES) measurements on $Bi_6O_6Se_2Cl_2$ single crystal, an optimal Cl-doping compound, we show that the key electronic features, including indirect band gap, high electron mobility, and carrier effective mass, are well preserved, while the 2D character of the system is sufficiently enhanced. First-principles calculations attribute this selective tunability to the unique spatial separation between conductive Bi-O layers and dopant Se/Cl layers, which enables independent optimization of interlayer coupling and intralayer transport. Remarkably, our theoretical calculations** **combined with experimental observations project a broad bandgap engineering across a wide range (0.8 ~ 3.5 eV) within the $Bi_2O_2Se$ system. Our findings not only expand the application potential of $Bi_2O_2Se$-based materials in optoelectronics, nanoelectronics, and fundamental research but also highlight spatial decoupling as a viable strategy for modifying detailed electronic structure of quasi-2D quantum systems.**

## Introduction:

Among the emerging 2D materials that possess exceptional properties, the quasi-2-dimensional (2D) bismuth oxyselenide ($Bi_2O_2Se$) has rapidly distinguished itself for its considerable potential in both industrial applications and fundamental physics research[1]. It features exceptional electronic properties with great environmental stability, including a moderate bandgap (~0.8 eV), ultrahigh intrinsic electron mobility (20,000 $cm^2V^{-1}s^{-1}$ at cryogenic temperatures), robustness of electronic properties against surface perturbations[2,3]. These properties, together with their inherent ability to form native high-permittivity (high-κ) oxide-insulator ($Bi_2O_2Se/Bi_2O_5Se$, mimicking $Si/SiO_2$)[4-6], make it a leading candidate as the post-silicon materials to sustain Moore's law. The construction of high-performance Fin-FET and gate-all-around FET (GAAFET) showcase its immediate potential for next-generation electronic devices[7,8]. Low-cost and broad-range photodetector with high responsivity and detectivity has also been constructed based on this material, illustrating its wide industrial application potential[9-11]. More importantly, the superior electronic properties of $Bi_2O_2Se$ also make it a versatile platform for fundamental physics research. The compensation of Rashba spin polarization in adjacent Bi-O layers (hidden Rashba effect) leads to the realization of unique quantum hall states (QHS), where only the even-integer states present[12-14]. Notably, this even-odd QHS transition could be further manipulated through field effect tuning[13], showcases its great potential in the investigation of fundamental quantum phenomena as well as future spintronics devices.

Despite its remarkable potential, an inherent limitation of the $Bi_2O_2Se$ crystal structure restricts its broader application. Unlike conventional van der Waals materials, where weakly interacting layers enable facile mechanical exfoliation, $Bi_2O_2Se$ is a quasi-2D material—its conductive $Bi_2O_2^{2+}$ layers are strongly bonded through bridging $Se^{2-}$ anions, forming robust ionic interactions along the out-of-plane direction (Fig. 1a). Consequently, obtaining high-quality ultrathin $Bi_2O_2Se$ requires specialized growth techniques such as molecular beam epitaxy (MBE)[12,15], limiting its accessibility for device fabrication. The introduction of van der Waals gaps into this system could sufficiently expand its research opportunities both in application and fundamental research. For instance, mechanical exfoliation and stacking would enable the heterostructure engineering study through proximity-induced phenomena and moiré potential engineering[16]. Reducing the interlayer coupling would improve electron performance while allowing better electrostatic control - crucial for making smaller transistors and quantum devices. In addition, enhancing 2D character would potentially accentuate quantum effects through confinement, where cleaner signals of quantum hall effects and stronger electron-electron correlations are expected. Previous research efforts brought up a feasible approach to address this issue through chlorine substitution of selenium; however, limited in the form of powder or polycrystals[17-21]. A complete substitution is also not favored since it transforms the system into a wide-bandgap insulator, BiOCl (~ 3.5 eV)[17], sacrificing the exceptional electronic properties that make $Bi_2O_2Se$ technologically valuable.

In this work, we report the synthesis and systematic electronic characterization of high-quality $Bi_6O_6Se_2Cl_2$ single crystals, an optimal chlorine doping that both enables mechanical exfoliation through the introduction of a van der Waals gap and preserves the basic electronic properties of the system. Combining the use of angle-resolved photoemission spectroscopy (ARPES) and transport measurement, we confirm that the key electronic features of $Bi_2O_2Se$ are maintained, while the two-dimensionality of the band structure is significantly enhanced. Key electronic parameters such as band gap size, intrinsic doping level, and effective mass of carriers are extracted, providing information for future research based on this system. Through theoretical analysis, we attribute this versatile tunability of the system to the unique crystal structure where the conductive Bi-O layers and dopant Se/Cl layers are spatially separated. This enables independent optimization of interlayer coupling to tune the band gap in a wide range, from 0.8 eV to 3.5 eV, while preserving the intralayer electronic band features. Our result not only opens up new avenue for further exploitation of $Bi_2O_2Se$-based materials, but also sets up an example of how targeted atom substitution can effectively modify the electronic structure of a quantum system.

**Result:**

Illustrated in Fig. 1a, the crystal structure of $Bi_2O_2Se$ features adjacent conductive Bi-O layers bridged by Se atoms through ionic bonding. Chlorine substitution turns a single Se layer into two Cl layers and establishes van der Waals connections instead. This modification results in $Bi_6O_6Se_2Cl_2$ adopting a 1:2 superlattice structure in the P4/nmm symmetry ($Bi_2O_2Se$-$Bi_2O_2Cl_2$-$Bi_2O_2Se$) with lattice parameters of a = b = 3.89 Å, c = 19.6 Å. The introduction of the van der Waals gap enables facile mechanical exfoliation with standard Scotch tape. As demonstrated in Fig. 1b, the optical image and atomic force microscopy (AFM) scanning reveal the exfoliated thin flakes with different thicknesses, where the step edges corresponding to 1- and 2-unit cells (2 nm and 4 nm, respectively) are evidenced. This demonstrates the feasibility of layer engineering and heterostructure fabrication based on this system. First-principles calculations predict the preservation of key electronic features with enhanced two-dimensionality (Fig. 1d). The system maintains the indirect band gap characteristic of $Bi_2O_2Se$, with the conduction band minimum located at Γ and valence band maximum at X. Significantly, the near-identical band dispersions along both the Γ (blue) and Z (red) planes, corresponding to different $k_z$ values in the Brillouin Zone (Fig. 1c), demonstrate markedly reduced out-of-plane dependence compared to pristine $Bi_2O_2Se$ [2]. This suggests a quasi-2D electronic behavior in the modified structure.

High-quality $Bi_6O_6Se_2Cl_2$ single crystals (Fig. 2a) used for this study were synthesized via the chemical vapor transport (CVT) method (see Methods for details), where the crystal structure is verified through the X-ray diffraction (XRD) measurement. Energy-dispersive X-ray spectroscopy (Fig. 2b and Fig. S1) reveals stoichiometric control and spatial homogeneity, showing a uniform Bi:Cl:Se ratio of 3:1:1 (±1%) across the samples, while X-ray photoelectron spectroscopy (Fig. 2c, Fig. S2) identifies the characteristic binding energies of each element, confirming their chemical states.

An overview of the general electronic behavior of this system can be obtained through electronic transport measurements. As shown in Fig. 2d, the longitudinal resistivity $\rho_{xx}$ of $Bi_6O_6Se_2Cl_2$ crystals exhibits metallic behavior, similar to that of the parent compound $Bi_2O_2Se$. This arises from the self-modulation doping effect of Se/Cl vacancies that populate the conduction band with additional electrons[22,23]. Consequently, low-temperature transport reflects heavy carrier doping rather than intrinsic thermal activation across the bandgap. Temperature-dependent resistivity $\rho_{xx}$ can be well fitted by $T^2$ ($\rho_{xx} = \rho_0 + a * T^2$, where $\rho_0$ is the resistivity at 0 K and $a$ is a constant) from 2 to 60 K (see Fig. 2d), which is a feature of Landau-Fermi liquid[24,25], suggesting a sole electron-electron scattering mechanism in the electrical transport at low temperature (<60 K). Hall measurement results (inset of Fig. 2d) corroborate this n-type conduction, showing the characteristic negative slope that yields an electron density of n=9.39×$10^{18}$ $cm^{-3}$ and a carrier mobility of μ= 395.46 $cm^2$ $V^{-1}$ $s^{-1}$ at 2 K (Fig. 2e). While this mobility is reduced compared to pristine $Bi_2O_2Se$ - likely due to enhanced electron scattering from Cl/Se antisite defects[18] in the current crystals (evidence discussed in Supplementary Information, Note 2 and Table S1-4) - it still surpasses most conventional 2D semiconductors[26] like $MoS_2$, highlighting the material's potential for device applications. At low temperatures, the magnetoresistance (MR) shows a characteristic weak localization (WL) cusp below 50 K (Fig. 2f), a signature of coherent electron interference effects that may arise from the existence of a strong spin-orbit interaction[27]. Quantitative fitting analysis (Fig. S3 and the result in Fig. 2g) of the magnetoconductance using the Hikami-Larkin-Nagaoka model (HLN)[28] reveals several intriguing aspects: the extracted phase coherence length ($L_\varphi$) of 43.86 nm at 2 K demonstrates well-preserved quantum effects, while the temperature-dependent suppression of this length follows expected trends for disordered systems[29,30]. Notably, $L_\varphi$ decreases with increasing temperature and follows a power law of $L_\varphi \sim T^{-0.42}$. This behavior, together with the prefactor value of α ~$10^2$ that falls between theoretical predictions for pure 2D and 3D systems[31-33], suggests our engineered superlattice structure achieves a quasi-2D electronic environment with enhanced quantum confinement.

To elucidate the detailed electronic band structure of the $Bi_6O_6Se_2Cl_2$ system, synchrotron-based high-resolution ARPES measurements were performed on the 001-cleavage plane of the crystals. From the 3D overview of the ARPES intensity (Fig. 3a), the conduction band minimum (CBM) is evidenced at the Γ point, while the valence band maximum (VBM) was found at the X point, confirming the indirect band gap nature of the system. The observed four-fold symmetry of the valence bands, as evident in the constant energy contours (Fig. 3b,c), is in line with the lattice symmetry. These general electronic features match closely with the parent compound $Bi_2O_2Se$, suggesting that chlorine substitution preserves the basic electronic structure. A detailed examination along the high-symmetry direction Γ-X-Γ (Fig. 3d) reveals an indirect band gap size of 1.10 ± 0.07 eV. Notably, the valence band edge exhibits slight broadening, which likely stems from the disorder effects induced by Se/Cl antisite defects[17,34,35], as these states are primarily derived from Se/Cl p-orbitals[17]. In contrast, the conduction band remains sharp. Zoomed-in analysis (Fig. 3e) reveals a self-doping level of ~ 0.4 eV, with two isotropic electron pockets observed. The effective mass of the electrons, obtained via parabolic fitting to the band dispersion (Fig. 3f), yields $m_e^* \sim 0.18 \pm 0.03\ m_0$ ($m_0$ is the mass of a free electron). In addition, two distinct hole pockets, one heavy and one light, coexist at the VBM. Their effective masses along the Γ-X direction are estimated to be $m_{h1}^* \sim -0.50 \pm 0.12\ m_0$, and $m_{h2}^* \sim -1.60 \pm 0.06\ m_0$, respectively. This coexistence of light and heavy hole bands makes the system a promising candidate for thermoelectric applications, as the light holes facilitate high electrical conductivity while the heavy holes help maintain a large Seebeck coefficient[21,36].

To evaluate the quasi-2D nature of the system, we further performed photon-energy-dependent ARPES measurements to probe the electronic structure across different $k_z$ values within the 3D Brillouin zone (BZ). Theoretical calculations predict a $k_z$-dependent evolution of the electron pocket (Fig. 1d), characterized by a varying energy separation between the two conduction bands between the Γ and Z planes. As illustrated in Fig. 4a, tuning the photon energy from 82 eV to 96 eV reveals that while the lower conduction band remains nearly stationary, the upper conduction band vanishes, as can be more clearly visualized in the constant energy contours of Fig. 4b. A wide-range photon energy scan covering multiple BZs (Fig. 4c) confirms the periodic reappearance of this upper conduction band, allowing for the precise identification of the high-symmetry Γ (blue) and Z (red) planes along the $k_z$ direction (see Supplementary Information Note I for detailed analysis). Despite the $k_z$-sensitivity of the upper conduction band, the primary electronic features exhibit remarkable similarity between the Γ and Z planes. This is demonstrated by the band dispersion cuts at $k_z = 0$, $\pi/3$, $2\pi/3$, and $\pi$ (Fig. 4d), which contrast sharply with the more dispersive nature of pristine $Bi_2O_2Se$[2]. A direct comparison of the dispersions and overlaid DFT calculations at the Γ and Z points (Fig. 4e) shows nearly identical features. These results signify successful electronic structure engineering, effectively enhancing the two-dimensionality of the system while preserving its fundamental electronic properties.

## Discussion:

The exceptional tunability of $Bi_2O_2Se$ through chlorine substitution originates from its unique layered structure and orbital characteristics. As illustrated in Fig. 5a, the system exhibits a natural electronic decoupling where conduction electrons, primarily derived from Bi p orbitals, remain strongly localized within the conductive Bi-O layers, while Se vacancies in the adjacent layers provide self-doping electrons that make the system intrinsically metallic[22]. This spatial separation creates a remarkable protection of the electrons from the disorder scatterings in the Se/Cl layers, resulting in a high carrier mobility of the system. The introduction of van der Waals gaps through Cl substitution maintains this scenario and further produces two complementary effects. It systematically reduces interlayer hybridization, enhancing two-dimensionality and thereby driving a monotonic widening of the bandgap (~0.8 eV to ~ 3.5eV), as indicated by the result of DFT calculation and experimental verification (Fig. 5b,c, and Table 1). On the other side, it preserves the pristine conduction band characteristics, such as band dispersion and effective mass (Fig. 5b-c). These two effects are confirmed with consistent experimental measurements and theoretical calculations (Fig. 5b), illustrating how spatial separation of functional sublattices allows for independent tuning of interlayer coupling while preserving desirable intralayer characteristics in this system. Detailed calculation showing the band structure evolution is presented in Fig. 5d, and the ARPES measurement on high-quality single crystal $Bi_4O_4SeCl_2$ (50% substitution) is presented in Fig. S4-5.

The unique electronic properties of chlorine-substituted $Bi_2O_2Se$ offer significant potential for diverse industrial applications. Specifically, the widely tunable bandgap and van der Waals (vdW) interactions enable the fabrication of photodetectors with high responsivity, high stability, continuous broadband tuning, simple manufacturing, and low cost[9-11]. The preserved high electron mobility facilitates high-performance, fast-response, nanoelectronic devices with low power consumption. Moreover, the coexistence of light and heavy hole bands, with suppressed phonon-mediated heat transport, positions this material family as a promising candidate for next-generation thermoelectric devices[21,36]. Meanwhile, multiple approaches should be applied to further optimize the quality of these Cl-substituted crystals. The current mobility limitations, evident from valence band smearing (Fig. 3d), primarily originate from Se-Cl antisite defects, which could be suppressed with optimization of growth parameters, specifically the tuning of Se/Cl partial pressures and the implementation of slower post-growth annealing cycles. Such enhancements would simultaneously benefit practical device performance and enable more precise investigations of quantum transport phenomena. Notably, this band structure engineering approach has broader implications beyond halogen substitution. Following the paradigm established in $TaS_2$ and related dichalcogenides family, where intercalation has unveiled rich phenomena including magnetic ordering, topological states, and unconventional superconductivity etc[37-39], we anticipate $Bi_2O_2Se$ to be a versatile platform for discovering novel quantum phenomena through similar structural modifications.

## Methods:

### Sample synthesis and basic characterizations

$Bi_6O_6Se_2Cl_2$ single crystals were grown by CVT with iodine ($I_2$, Alfa-Aesar, 99.9985%) as the transport agent. We first synthesized $Bi_6O_6Se_2Cl_2$ polycrystalline powders by the direct stoichiometric solid-state reaction of high pure $Bi_2O_3$ powder (Sinopharm Chemical Reagent Co., Ltd, 99.999%), BiOCl powder (Alfa Aesar, 99.99%), bismuth powder (Alfa Aesar, 99.999%) and selenium powder (Alfa Aesar, 99.999%) with a 2:6:2:3 molar ratio in a fused silica tube sealed under vacuum ($P \sim 4\times10^{-6}$ Torr) at about 850°C for several days. Then the mixture of $Bi_6O_6Se_2Cl_2$ and $I_2$ (about 10 mg/mL) powders were loaded into the sealed evacuated quartz tube ($P \sim 4\times10^{-6}$ Torr), and then put into a two-zone furnace to grow crystals. The raw materials were placed at the hot zone of the ampoule to maintain the crystal growth and the empty end of the quartz tube was maintained at a lower temperature to grow crystals. The millimeter-sized crystals with metallic luster were obtained successfully after growing at a temperature profile of 750–650 °C for over 14 days. To determine the surface orientation of the $Bi_6O_6Se_2Cl_2$ single crystals, X-ray diffraction (XRD) measurements were conducted on an X-ray diffractometer (Ultima III Rigaku) using Cu-$K_\alpha$ radiation with a 2θ range of 5~85°. The composition of single crystals was characterized by an energy-dispersive spectroscopy (EDS) spectrometer equipped with a scanning electron microscope (FEI Inc., Quanta). Longitudinal resistivity $\rho_{xx}$ and Hall (transversal) resistivity $\rho_{xy}$ were measured using a standard four-probe configuration on a 9 T Quantum Design Physical Property Measurement System (PPMS).

**Angle-resolved photoemission spectroscopy (ARPES)**

ARPES measurements were performed at BL03U of the Shanghai Synchrotron Radiation Facility (SSRF)[40]. The data were recorded using a DA30L analyzer with the total convolved energy and angle resolutions of 10 meV and 0.2°, respectively. The samples were cleaved in situ on the (001) cleavage surface and the measurements were carried out under ultrahigh vacuum of better than $1.3\times10^{-10}$ mbar at the temperature of 12 K.

**Theoretical calculation**

The band structure was calculated by the density functional theory in the generalized gradient approximation implemented in the Vienna ab-initio simulation package (VASP)[41]. The projected augmented wave method and the Perdew-Burke-Ernzerhof exchange correlation are used in the calculations within the generalized gradient approximation (GGA)[42]. The plane-wave cutoff energy was set to 520 eV, with the Brillouin zone sampled by a 9×9×9, 6×6×6, 8 × 8 × 2, 10 × 10 × 1, and 9 × 9 × 6 k-grid for $Bi_2O_2Se$, $Bi_4O_4SeCl_2$, $Bi_6O_6Se_2Cl_2$, $Bi_6O_6SeCl_4$/$Bi_8O_8Se_3Cl_2$, and BiOCl, respectively. Except for $Bi_2O_2Se$, the optB86b-vdW correlation functional has been used in the calculation. To obtain accurate band gaps, we also employ the Heyd-Scuseria-Ernzerhof (HSE06) hybrid functional in the band structure calculations[43,44].

**Acknowledgement**

We thank the financial support from the National Natural Science Foundation of China (Grant no. 52331008, 12474044, 92463304, 52272002, and 52572008), the National Key Research and Development Program of China (Grant no. 2024YFA1410400), and the Natural Science Foundation of Jiangsu Province (Grant No. BK20253012, No. BK20242013, and No. BK20241805). The numerical calculations in this paper have been done on the computing facilities in the High-Performance Computing Center (HPCC) of Nanjing University.

## Competing Interests

The authors declare no competing interests.

## Main text figure captions:

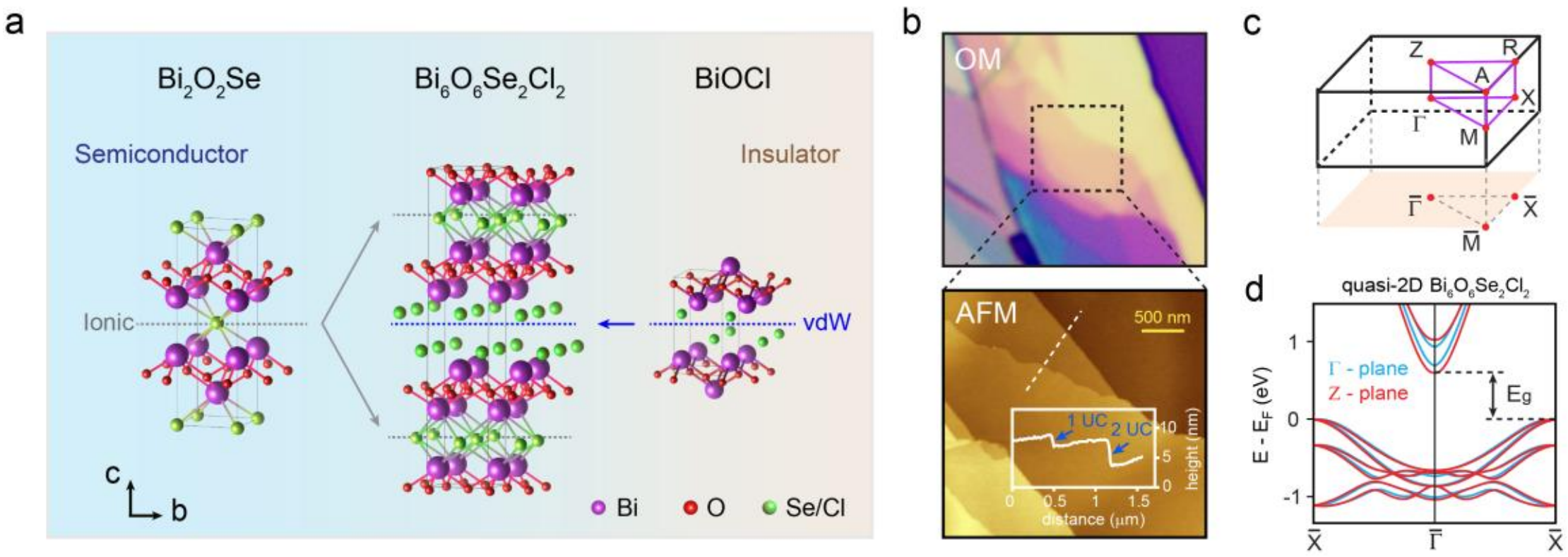

**Fig. 1| Crystal structure and quasi-2D properties of $Bi_6O_6Se_2Cl_2$. a**, Comparison of the crystal structures of $Bi_2O_2Se$, $Bi_6O_6Se_2Cl_2$, and BiOCl. The substitution of Se by Cl atoms introduces van der Waals (vdW) gaps into the lattice. **b,** Optical microscope (OM) and atomic force microscope (AFM) images of mechanical exfoliated thin flake of $Bi_6O_6Se_2Cl_2$. Height profiles (inset) confirm step heights corresponding to 1- and 2-unit cells. **c,** Brillouin zone and its projection onto the (001) surface with high-symmetry points labeled. **d,** Calculated band dispersions along the $\Gamma$ and $Z$ planes. The nearly identical dispersions indicate the quasi-2D electronic nature of the system.

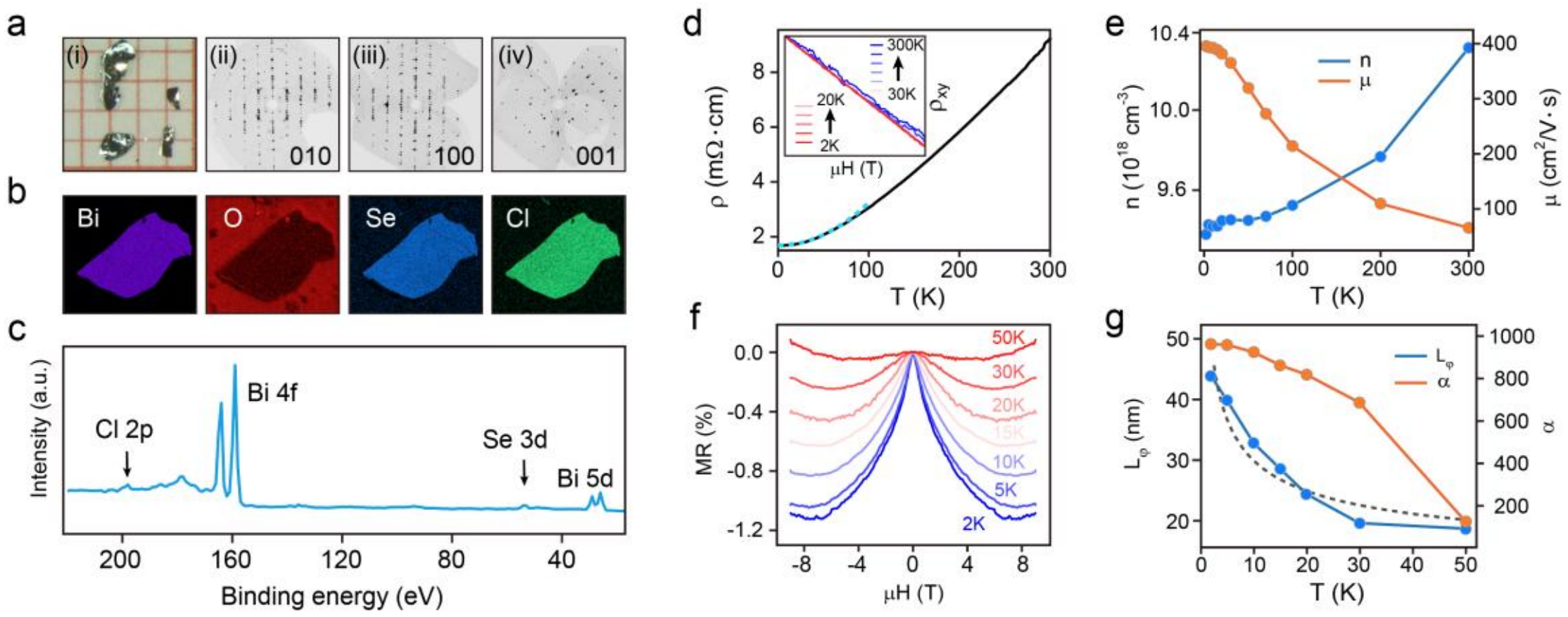

**Fig. 2| Crystal quality and transport measurements of $Bi_6O_6Se_2Cl_2$. a,** (i) Image of high-quality $Bi_6O_6Se_2Cl_2$ single crystal. (ii-iv) XRD patterns along (010), (100), and (001) surfaces. **b,** EDX image of exfoliated $Bi_6O_6Se_2Cl_2$ flakes showing spatial uniformity of the elements. **c,** Core-level photoemission spectrum, showing the characteristic peaks of Bi, Se, and Cl core levels. a.u., arbitrary units. **d,** Longitudinal resistivity $\rho_{xx}$ as a function of temperature of $Bi_6O_6Se_2Cl_2$ single crystal. The dotted line is the fitting at low temperature using $\rho_{xx} = \rho_0 + AT^2$. Inset: Hall resistivity $\rho_{xy}$ as functions of out-of-plane magnetic field with temperatures from 2 to 300 K. **e,** Extracted carrier densities $n$ and carrier mobility $\mu$ as a function of temperature. **f,** Magnetoresistance (MR) of out-of-plane magnetic field with varying temperatures. **g,** Extracted phase coherence length $L_\varphi$ and prefactor α as a function of temperature under a field limit of 3T. The dotted line is the power-law fitting of the coherence length, which yields $L_\varphi \sim T^{-0.42}$.

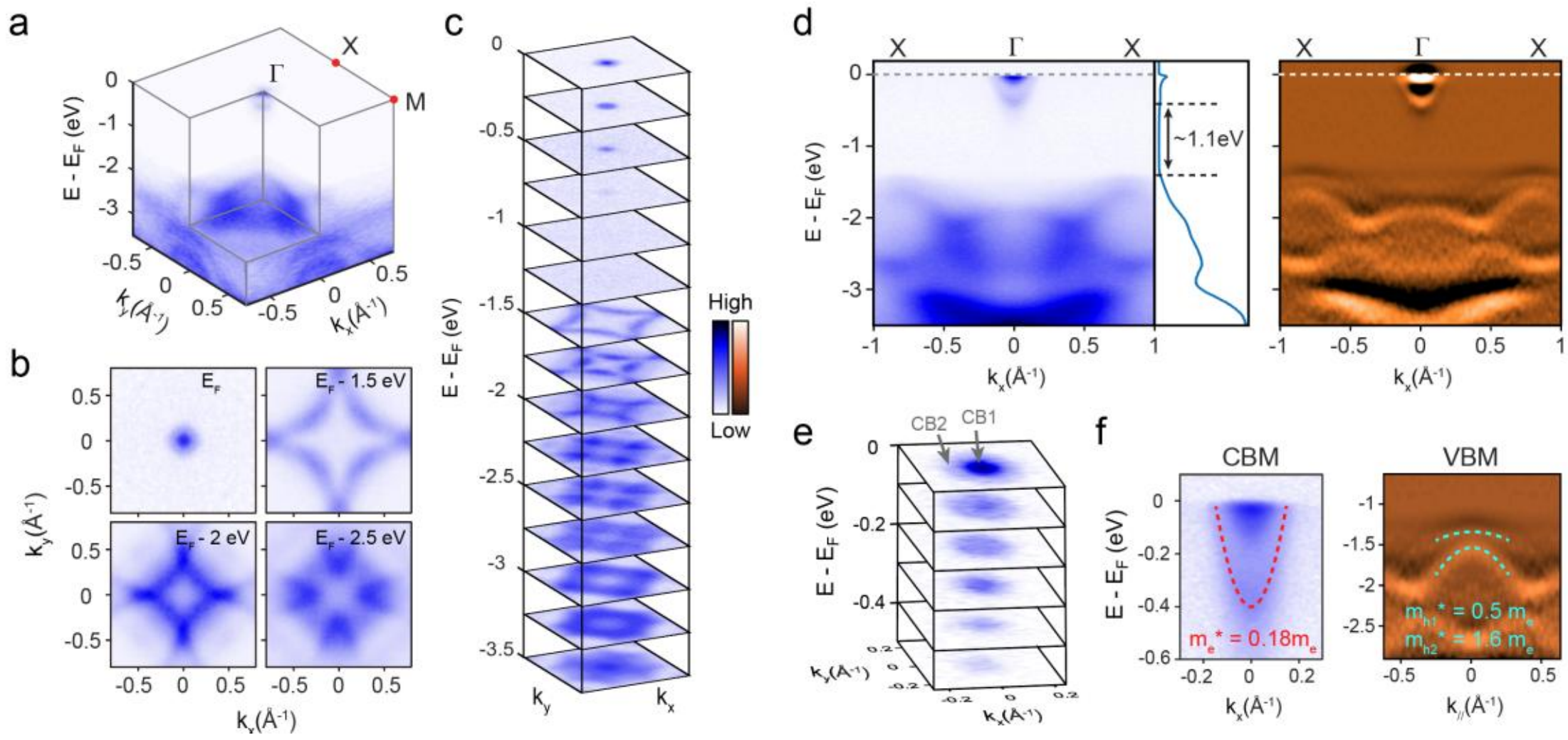

**Fig. 3| General electronic band structure of $Bi_6O_6Se_2Cl_2$. a,** 3D plot of the photoemission intensity measured on $Bi_6O_6Se_2Cl_2$ single crystal. Conduction band bottom and valence band top is evidenced at Γ and X point respectively. **b,** Constant energy contours of the photoemission intensity on Fermi level ($E_F$) and at binding energies of -1.5, -2, and -2.5 eV, respectively. **c,** Stacking plots of constant energy contours showing the complete band structures. **d,** Band dispersion cut and its second derivative plot showing the band structure along $\mathrm{X}-\Gamma-\mathrm{X}$ direction. The extracted density of state (DOS) is plotted on the side, illustrating an indirect band gap of ~1.1 eV. **e,** Stacking plot zoomed in to the two conduction bands (CB), with detailed dispersion shown in f. **f,** Fitting of the conduction band minimum (CBM) and valence band maximum (VBM) along $\Gamma-\mathrm{X}$ direction.

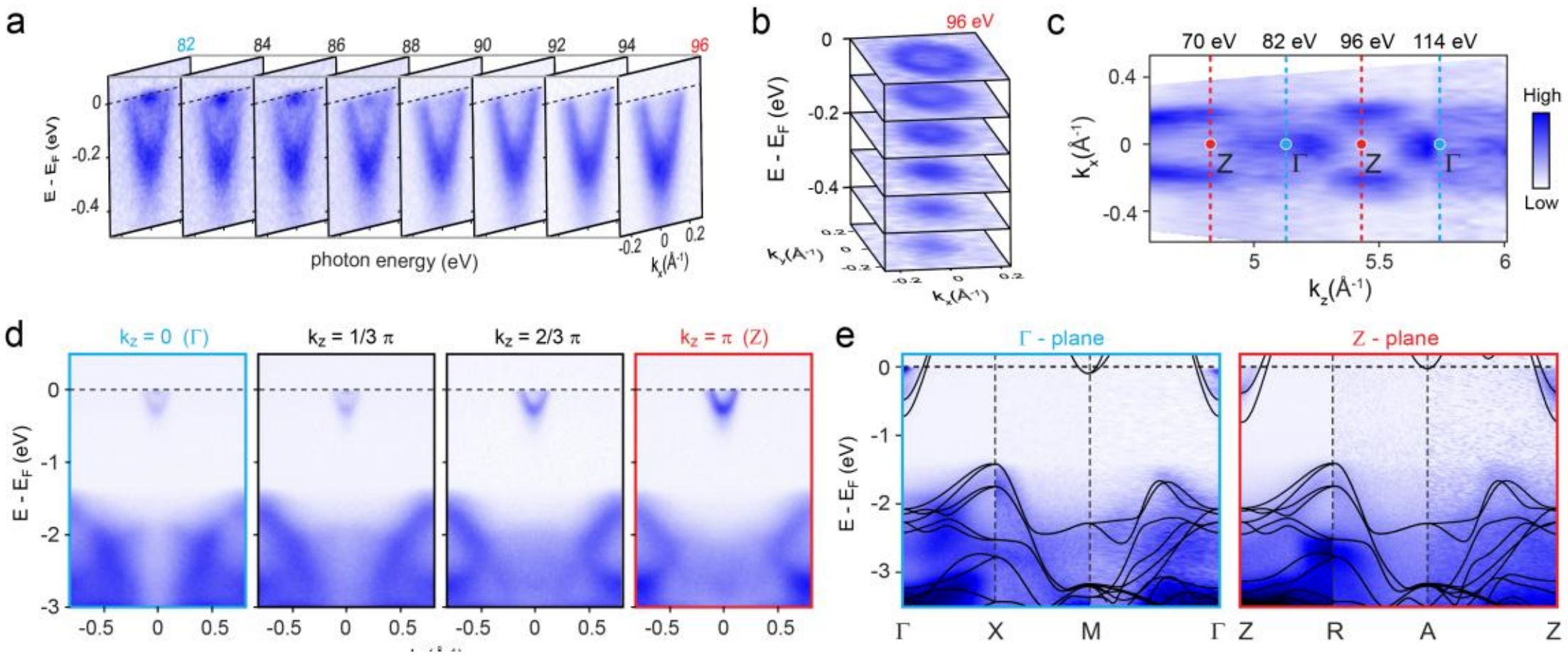

**Fig. 4| Detailed quasi-2D electronic features in full Brillouin Zone. a,** Photoemission intensity of electron pockets with varying photon energies from 82 eV (Γ-plane) to 96 eV (Z-plane), showing its evolution with out-of-plane momentum ($k_z$). **b,** Stacking plot of the electron pocket on the Z-plane. **c,** Fermi surface of $k_x$-$k_z$ plane, showing the periodic evolution of two conduction bands. High symmetry positions Γ and Z are labeled. **d,** High statistic cuts of the band dispersion along $\mathrm{X}-\Gamma-\mathrm{X}$ direction at different $k_z$ value of $0,\ \pi/3,\ 2\pi/3,\ \pi$ respectively, showing almost identical features. **e,** Detailed band dispersion along high symmetry directions on both the Γ-plane and Z-plane with DFT calculation overlaid, indicative of the quasi-2D nature of the system.

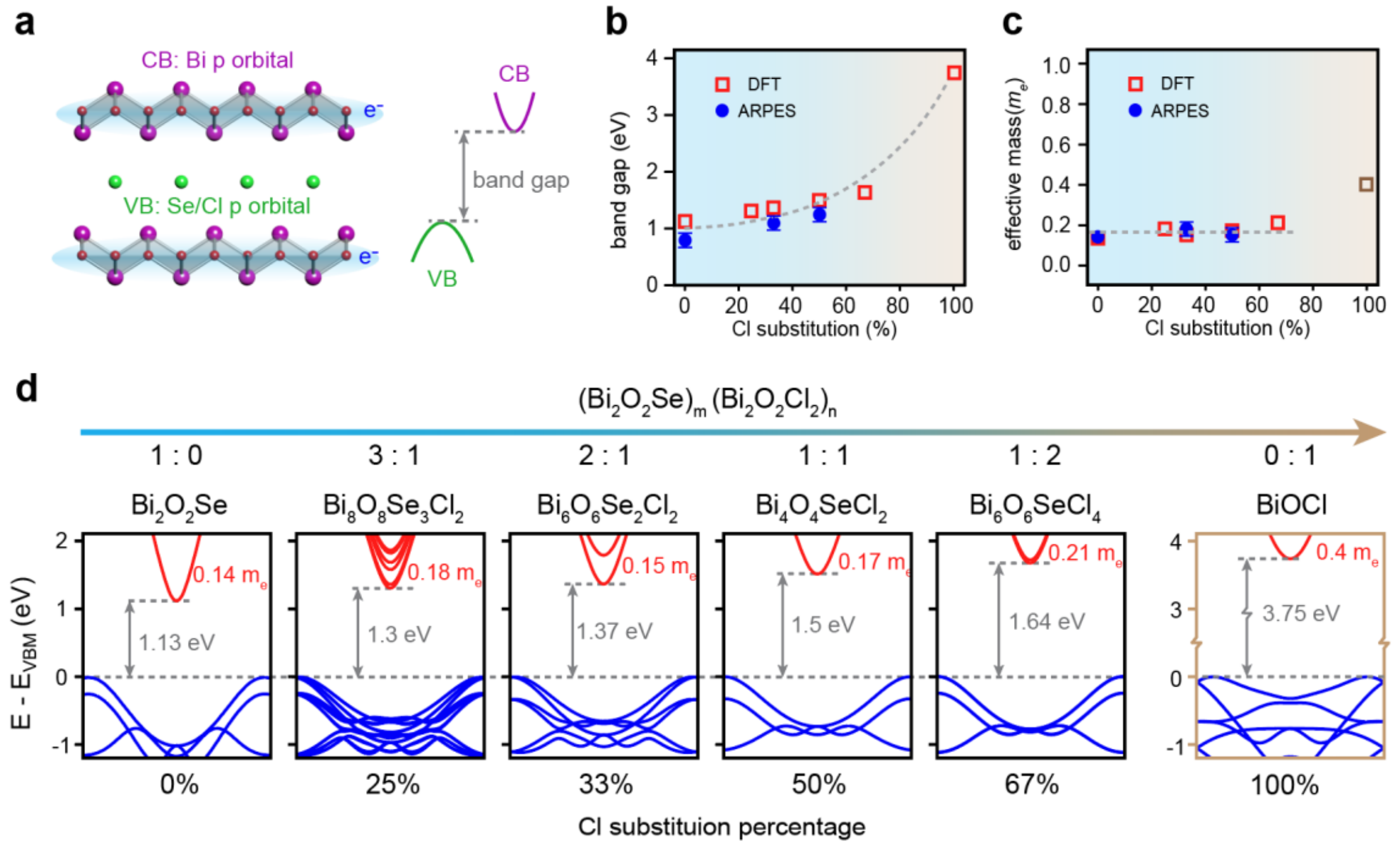

**Fig. 5| Rational band structure engineering with Cl substitution in $Bi_2O_2Se$ family. a,** Schematic showing the spatial separation of the conductive Bi-O layer and the dopant Se/Cl layer. The conduction band mainly consists of Bi p orbitals, while the valence bands are contributed by Se/Cl p orbitals. **b-c,** Evolution of the indirect band gap size and effective mass of the conduction band (derived from both DFT calculation and ARPES experiments, summarized data in Table 1) as a function of Cl substitution over Se. With the weakening of the interlayer interaction in Se/Cl layer, the band gap size of the system increases monotonically, while the pristine conduction band characteristics from Bi-O layer (e.g. effective mass of electron pocket) are preserved. Data for $Bi_2O_2Se$ is from Ref [2]. **d,** Detailed band dispersion along $X-\Gamma-X$ direction as a function of Cl substitution, showing systematic evolution.

| | $m_e^*$ | $m_h^*$ | band gap |
|---|---|---|---|
| **$Bi_2O_2Se$ (Cl 0%)** | $(0.14 \pm 0.02)\ m_0$ | $(-2.41 \pm 0.02)\ m_0$ | $(0.80 \pm 0.05)$ eV |
| **$Bi_6O_6Se_2Cl_2$ (Cl 33%)** | $(0.18 \pm 0.03)\ m_0$ | $(-0.50 \pm 0.12)\ m_0$ (h1)<br>$(-1.60 \pm 0.06)\ m_0$ (h2) | $(1.10 \pm 0.07)$ eV |
| **$Bi_4O_4SeCl_2$ (Cl 50%)** | $(0.16 \pm 0.02)\ m_0$ | $(-1.2 \pm 0.1)\ m_0$ | $(1.25 \pm 0.07)$ eV |

**Table 1: Extracted band parameters from ARPES spectra. $m_e^*/m_h^*$:** effective mass of electron/hole pockets in the unit of free electron $m_0$. Data of $Bi_2O_2Se$ is adapted from Ref. 2.